\pdfoutput=1
 
\documentclass[aps,prl,preprint,groupedaddress,amsmath,amssymb]{revtex4-1}

\usepackage{graphicx}
\usepackage{dcolumn}% Align table columns on decimal point
\usepackage{bm}% bold math

\begin{document}

\title{Beam-beam effects investigation and parameters optimization for a circular e+e- collider TLEP to study the Higgs boson\thanks{Work is supported by the Ministry of Education and Science of the Russian Federation}}

\author{ A.~Bogomyagkov }
  \altaffiliation[Also at ]{Novosibirsk State University, Novosibirsk 630090, Russia}
  \email{A.V.Bogomyagkov@inp.nsk.su}
\author{ E.~Levichev }
  \altaffiliation[Also at ]{Novosibirsk State Technical University, Novosibirsk 630073, Russia}
\author{ D.~Shatilov }
\affiliation{Budker Institute of Nuclear Physics SB RAS, Novosibirsk 630090, Russia}

\date{\today}

\begin{abstract}
Several proposals exist for future circular electron-positron colliders designed for precise measurements of the Higgs boson characteristics and electroweak processes.
At very high energies, synchrotron radiation of the particles in a strong electromagnetic field of the oncoming bunch ({\it beamstrahlung}) becomes extremely important, because of degradation of the beam lifetime and luminosity. We present theoretical calculations of beamstrahlung (including the beam lifetime reduction and the energy spread increase) which are benchmarked against quasi strong-strong computer simulation. Calculation results are used to optimize TLEP project (CERN).
\end{abstract}



\maketitle

\section{Introduction}
Design study has commenced of high luminosity $e^+e^-$ collider TLEP for precise measurements of the Higgs boson properties and other experiments at the electroweak scale at CERN. TLEP will be capable to collide beams in wide center-of-mass energy range from 90 to 350 GeV (with an option up to 500 GeV) with luminosity higher than $5\cdot10^{34} cm^{-2} s^{-1}$ \cite{HighLuminosityCollider}. 

As mentioned in \cite{Augustin:1978ah}, a key issue that limits luminosity and beam lifetime in circular electron-positron colliders with high energy is beamstrahlung, i.e., synchrotron radiation of a lepton deflected by the collective electromagnetic field of the opposite bunch. Because of this radiation, colliding particles of TLEP at high energy could lose so much energy that they are taken out of the momentum acceptance of accelerator (beam lifetime limitation due to the {\it single} beamstrahlung). In the beginning of 2013, V.~Telnov estimated lifetime considering single beamstrahlung \cite{PhysRevLett.110.114801}, and set of TLEP parameters using V.~Telnov's formula was given in \cite{TLEPparametersCERN}. For TLEP at low energies, energy loss because of beamstrahlung is not large enough to kick the particles immediately out of the momentum acceptance; however {\it multiple} beamstrahlung increases beam energy spread and bunch length, reducing luminosity owing to the hour glass effect.

We present an analytical approach to calculate the beam lifetime limitation caused by the single beamstrahlung as well as the energy spread and bunch length increase due to the multiple beamstrahlung. Results of the theoretical predictions are compared with weak-strong beam-beam tracking code Lifetrac  \cite{LifeTrackShatilov}, in which the effect of beamstrahlung was introduced. Set of new parameters of TLEP with higher luminosity or/and better lifetime is presented for further studies. We considered head-on and crab waist \cite{CrabWaistPanta} collisions schemes.

\section{Analytical calculations}
\subsection{Beam-beam}
The potential of incoming beam is written as
\begin{widetext}
\begin{equation}
\label{eq:3dpotential}
U(x,y,s,z)=-\frac{2 N_p r_e}{\sqrt{\pi}}
\int_0^\infty
\frac{
\exp\left[\displaystyle -\frac{(x+s\, 2\theta)^2}{2\sigma_x^2+q}-\frac{y^2}{2\sigma_y^2+q}-\frac{\gamma^2(2s-z)^2}{2\gamma^2\sigma_s^2+q}\right]
}
{\sqrt{(2\sigma_x^2+q)(2\sigma_y^2+q)(2\gamma^2\sigma_s^2+q)}}dq\,,
\end{equation}
\end{widetext}
where $r_e$ -- classical electron radius, $\gamma$ -- Lorentz factor, $N_p$ -- amount of particles, $\sigma_{x,y,s}$ -- horizontal, vertical and longitudinal beam sizes, $2\theta$ -- crossing angle,${x,y,s}$ -- horizontal, vertical and longitudinal coordinates, $z=s-ct$ -- particle's position with respect to the center of the bunch describing synchrotron oscillations. For simplicity, we will neglect particle's synchrotron oscillations therefore $z=0$. \\
Equations of motion are written as:
\begin{widetext}
\begin{eqnarray}
\label{eq:F-y-0}
y'' &= & -\frac{\partial U}{\partial y}=-\frac{4 N_p r_e}{\sqrt{\pi}}y\int_0^\infty
\frac{
\exp\left[\displaystyle -\frac{(x+s\, 2\theta)^2}{2\sigma_x^2+q}-\frac{y^2}{2\sigma_y^2+q}-\frac{\gamma^2(2s)^2}{2\gamma^2\sigma_s^2+q}\right]
}
{\sqrt{(2\sigma_x^2+q)(2\sigma_y^2+q)^3(2\gamma^2\sigma_s^2+q)}}dq\,, \\
\label{eq:F-x-0}
x'' & = & -\frac{\partial U}{\partial x}=-\frac{4 N_p r_e}{\sqrt{\pi}}x\int_0^\infty
\frac{
\exp\left[\displaystyle -\frac{(x+s\, 2\theta)^2}{2\sigma_x^2+q}-\frac{y^2}{2\sigma_y^2+q}-\frac{\gamma^2(2s)^2}{2\gamma^2\sigma_s^2+q}\right]
}
{\sqrt{(2\sigma_x^2+q)^3(2\sigma_y^2+q)(2\gamma^2\sigma_s^2+q)}}dq\,.
\end{eqnarray}
\end{widetext}

In order to calculate effective interaction length $L$ and mean bending radius $\rho_{x,y}$ in hard edge approximation we will neglect $\sigma_{x,y}$ dependence on $s$ and find expected value of vertical $\Delta y'$ and horizontal $\Delta x'$ kicks. After calculations we obtained
\begin{eqnarray}
\label{eq:Kick-y}
\left<\left| \Delta y'\right|\right>_{y, x+s\,2\theta=0} & \approx & \sqrt{\frac{\pi}{2}}\frac{N_p r_e}{\gamma \sigma_x\sqrt{1+\phi^2}}\,, \\
\label{eq:Kick-x}
\left<\left| \Delta x'\right|\right>_{x, y=0} & \approx &  \frac{\log{ \left( \frac{\sqrt{2}+1}{\sqrt{2}-1}\right)}}{\sqrt{\pi}}\frac{N_p r_e}{\gamma \sigma_x\sqrt{1+\phi^2}}\,,
\end{eqnarray}
where $\phi=\sigma_s \theta/\sigma_x$ is Piwinski parameter, $\left<\right>$ means expected value with respect to the first coordinate in subindex while other one satisfies condition in subindex. Inverse bending radius in corresponding plane is calculated as
\begin{eqnarray}
\label{eq:F-y-1}
\frac{1}{\rho_y}&=&\left<\left|y''\right|\right>_{y, x+s\,2\theta=0}  \approx  \frac{N_p r_e}{\gamma \sigma_s\sigma_x}\,, \\
\label{eq:F-x-1}
\frac{1}{\rho_x}&=&\left<\left|x''\right|\right>_{x,y=0}  \approx \frac{\sqrt{2}}{\pi} \log{ \left( \frac{\sqrt{2}+1}{\sqrt{2}-1}\right)} \frac{N_p r_e}{\gamma \sigma_s\sigma_x}\,.
\end{eqnarray}
Finally effective interaction length in each plane is
\begin{equation}
\label{eq:InteractionLength-0}
L=L_x=L_y=\frac{\left<\left| \Delta y'\right|\right>_{y, x+s\,2\theta=0}}{\left<\left|y''\right|\right>_{y, x+s\,2\theta=0}}=\frac{\left<\left| \Delta x'\right|\right>_{x, y=0}}{\left<\left|x''\right|\right>_{x,y=0}}=\sqrt{\frac{\pi}{2}}\frac{\sigma_s}{\sqrt{1+\phi^2}}\,.
\end{equation}
Since $\log{ \left( \frac{\sqrt{2}+1}{\sqrt{2}-1}\right)}\sqrt{2}/\pi\approx 0.8$ (from (\ref{eq:F-x-1})) we will count horizontal and vertical bending radii as equal
\begin{equation}
\label{eq:BendingRadius-0}
\frac{1}{\rho_x}\approx\frac{1}{\rho_y}\approx \frac{N_p r_e}{\gamma \sigma_s\sigma_x}\,.
\end{equation}

\subsection{Beamstrahlung}
Following approach given in \cite{PhysRevLett.110.114801} amount of emitted photons and beam lifetime are given in equations (\ref{eq:AmountPhotons-0}) and (\ref{eq:LifeTime-0}). The only difference is that we do not make an assumption of $10\%$ of the particles experiencing the maximum field, but use average values calculated in previous paragraph.
\begin{equation}
\label{eq:AmountPhotons-0}
N(u>\eta E_0)=\frac{3}{4\sqrt{\pi}} \sqrt{\frac{\alpha r_e}{\eta}}\exp\left(-\frac{2}{3}\frac{\eta\alpha\rho}{r_e\gamma^2}\right)\frac{L\gamma^2}{\rho^{3/2}}\,,
\end{equation}
\begin{equation}
\label{eq:LifeTime-0}
\tau_{bs}=\frac{1}{f_0 N}=\frac{1}{f_0}\frac{4\sqrt{\pi}}{3} \sqrt{\frac{\eta}{\alpha r_e}}\exp\left(\frac{2}{3}\frac{\eta\alpha\rho}{r_e\gamma^2}\right)\frac{\rho^{3/2}}{L\gamma^2}\,,
\end{equation}
where $\alpha$ -- fine-structure constant, $N_{ip}$ -- number of IPs.

The difference from V.~Telnov's calculations is in estimation of interaction length $L$ which is given in equation (\ref{eq:InteractionLength-0})
\begin{equation}
\label{eq:InteractionLength-1}
L=\sqrt{\frac{\pi}{2}}\frac{\sigma_s}{\sqrt{1+\phi^2}}\quad\left(L_{Telnov}=\frac{\sigma_s}{2}\right)\,,
\end{equation}
and in expression for the total bending radius $\rho$ given in equation (\ref{eq:BendingRadius-2}) ($\rho_x$ and $\rho_y$ from (\ref{eq:BendingRadius-0}))
\begin{equation}
\label{eq:BendingRadius-2}
\frac{1}{\rho} =\sqrt{\frac{1}{\rho_x^2}+\frac{1}{\rho_y^2}}\approx\frac{N_p r_e}{\gamma \sigma_x\sigma_s}\sqrt{2} 
\quad\left(\frac{1}{\rho_{Telnov}}\approx\frac{N_p r_e}{\gamma \sigma_x\sigma_s}2\right) \,.
\end{equation}
The radiation integrals \cite{Helm:1973xn} are modified according to 
\begin{eqnarray}
\label{eq:RadIntegrals-1}
\Delta I_2 &= & \left(\frac{L}{\rho_x^2}+\frac{L}{\rho_y^2}\right)N_{ip}\,, \\
\label{eq:RadIntegrals-2}
\Delta I_3 &= & \frac{L}{\rho^3}N_{ip}\,,
\end{eqnarray}
where $N_{ip}$ is a number of interaction points.\\
Hence, we obtain expression for the beam lifetime
\begin{equation}
\label{eq:LifeTimeOurs}
\tau_{bs}=\frac{1}{f_0}\frac{4\sqrt{\pi}}{3} \sqrt{\frac{\eta}{\alpha r_e}}\exp\left(\frac{2}{3}\frac{\eta\alpha}{r_e\gamma^2}\times\frac{\gamma \sigma_x \sigma_s}{{\bf \sqrt{2}}r_eN_p}\right)\frac{{\bf \sqrt{2}}}{{\bf \sqrt{\pi}}\sigma_s\gamma^2}\left(\frac{\gamma \sigma_x \sigma_s}{{\bf \sqrt{2}}r_eN_p}\right)^{3/2}\,,
\end{equation}
where bold symbols are showing the difference from the expression given by V.I.Telnov \cite{PhysRevLett.110.114801}
\begin{equation}
\label{eq:LifeTimeTelnov}
\tau_{bs}=\frac{{\bf 10}}{f_0}\frac{4\sqrt{\pi}}{3} \sqrt{\frac{\eta}{\alpha r_e}}\exp\left(\frac{2}{3}\frac{\eta\alpha}{r_e\gamma^2}\times\frac{\gamma \sigma_x \sigma_s}{{\bf 2}r_eN_p}\right)\frac{{\bf 2}}{\sigma_s\gamma^2}\left(\frac{\gamma \sigma_x \sigma_s}{{\bf 2}r_eN_p}\right)^{3/2}\,.
\end{equation}

\section{The model used in beam-beam simulations}
To track a test particle through IP, the opposite (strong) bunch is represented by a number of thin slices. The trajectory's bending radius for each slice can be estimated as
\begin{equation}
\label{eq:BendingRadius-3}
\rho\approx\frac{\Delta s}{\Delta p/p}\,,
\end{equation}
where $\Delta s$ is effective slice width, $\Delta p$ –- the transverse component of beam-beam kick. Radiation spectrum corresponds to normal synchrotron radiation from a bending magnet if the following condition is satisfied
\begin{equation}
\label{eq:LongMagnet}
\left(\frac{\Delta p}{p}\right)_{total}\gg\frac{1}{\gamma}\,.
\end{equation}
Here $\left(\Delta p/p\right)_{total}$ stays for the entire bunch (not a slice!) and can be estimated as $4\pi\xi\sigma'\sim 10^{-3\div-4}$. The given condition is always satisfied at the large energies (e.g. TLEP, $\gamma\geq10^5$). The critical energy of radiation $u_c$ (in units of mean beam energy $E_0=\gamma_0 mc^2$) is
\begin{equation}
\label{eq:CriticalEnergy-0}
\frac{u_c}{E_0}=\frac{3}{2}\gamma_0^2\left(1+\frac{\delta_E}{E_0}\right)^3\frac{r_e}{\alpha\rho}\,,
\end{equation}
where $\delta_E$ is particle's energy deviation.
Hereinafter, the energy of emitted photons is always normalized with respect to critical energy $u_c$. The spectrum density of radiation is
\begin{equation}
\label{eq:spectrum-0}
\frac{d}{dt}n(u/u_c)=\frac{\sqrt{3}}{2\pi}\alpha\gamma\frac{c}{\rho}\int_{u/u_c}^\infty K_{5/3}(x)dx\, d\!\left(\frac{u}{u_c}\right)\,.
\end{equation}
Note that at relatively small energies, where (\ref{eq:LongMagnet}) becomes invalid, $u_c$ drops significantly and we can neglect the whole effect of beamstrahlung, therefore there is no need to be concerned about the spectrum. Taking into account the time of interaction: $\Delta t=\Delta s/c$, we obtain the (average) number of emitted photons in a small interval of spectrum:
\begin{equation}
\label{eq:spectrum-1}
\Delta n(u/u_c)=\frac{\sqrt{3}}{2\pi}\alpha\gamma\frac{\Delta p}{p}\int_{u/u_c}^\infty K_{5/3}(x)dx\, \Delta\!\left(\frac{u}{u_c}\right)\,.
\end{equation}
The actual number of emitted photons is given by Poisson distribution. For tracking purposes we replace the continuous spectrum by a sequence of discrete lines, from 0.01 to 20 with a step of 0.01 (all in units of $u_c$) -- 2000 in total. The lower and upper limits were chosen from the condition that the radiation power outside the borders is negligible. The step between the lines is small enough to adequately represent the spectrum. Since the critical energy $u_c$ also depends on the actual particle's trajectory, the overall spectrum of emitted photons in simulations will be continuous regardless of being discrete in units of $u_c$. Considering randomness (and rather low probability) of photon emission in any given interval of $\Delta(u/u_c)$, we conclude that our spectrum simplifications will not affect the final results.

We have $\Delta(u/u_c) = 0.01$ in (\ref{eq:spectrum-1}) and our lines correspond to spectrum intervals of $0.005\div0.015$ (1st), $0.015\div0.025$ (2nd), etc. The integrals of $K_{5/3}(x)$ were calculated once and written in a static table for all 2000 points. The sum of all these values is responsible for the total (average) number of emitted photons
\begin{equation}
\label{eq:MeanNumberPhotons}
\bar{n}=\frac{\sqrt{3}}{200\pi}\alpha\gamma\frac{\Delta p}{p}\sum_{m=1}^{2000}\int_{m/100}^\infty K_{5/3}(x)dx\,.
\end{equation}

The overall simulation algorithm is as follows. First, $\Delta p/p$ is calculated for each particle after passing a single slice of the opposite bunch. Second, the $u_c$ is calculated from (\ref{eq:BendingRadius-3}) and (\ref{eq:CriticalEnergy-0}), and  $\bar{n}$ -- from (\ref{eq:MeanNumberPhotons}). Then, the actual number of emitted photons $N_{ph}$ (which can be zero) is obtained from the Poisson distribution with parameter $\bar{n}$, using random number uniformly distributed in the interval of $[0, 1]$. The energy of each particular photon is defined according to the relative probabilities (which are proportional to integrals of $K_{5/3}(x)$) for different spectrum lines, using another random number in the interval of $[0, 1]$. In total, the random number generator is called $N_{ph}+1$ times for each particle-slice interaction.

It is noteworthy, beamstrahlung simulations are not affected by the number of slices $N_{sl}$ -- if it is large enough to correctly represent the opposite bunch. For example, further increase of $N_{sl}$ leads to proportional decrease of both $\Delta s$ and $\Delta p/p$, while $\rho$ and $u_c$ remain unchanged. The total number of emitted photons also does not change: $\bar{n}$ for each slice decreases with $\Delta p/p$, but it is compensated by $N_{sl}$ increase.

TLEP has 4 interaction points (TABLE \ref{tbl:MainParametersCERN}), therefore lattice is assumed to possess 4-fold symmetry and we chose fractional betatron phase advances between IPs (0.53,0.57).

Beamstrahlung influence makes the bunch longer, and also depends on the bunch length. Therefore simulation was performed by quasi strong-strong method, where in the several repeated iterations the weak and strong bunches exchanged their roles and the length of the weak bunch was assigned geometric mean of strong and weak bunches. The equilibrium of the bunch length was found. Simulation was performed by weak-strong beam-beam tracking code Lifetrac \cite{LifeTrackShatilov}.

\section{Comparison of our results with previous}
\begin{figure}
\includegraphics[width=12cm,angle=0,trim=19mm 89.0mm 19mm 88mm, clip]{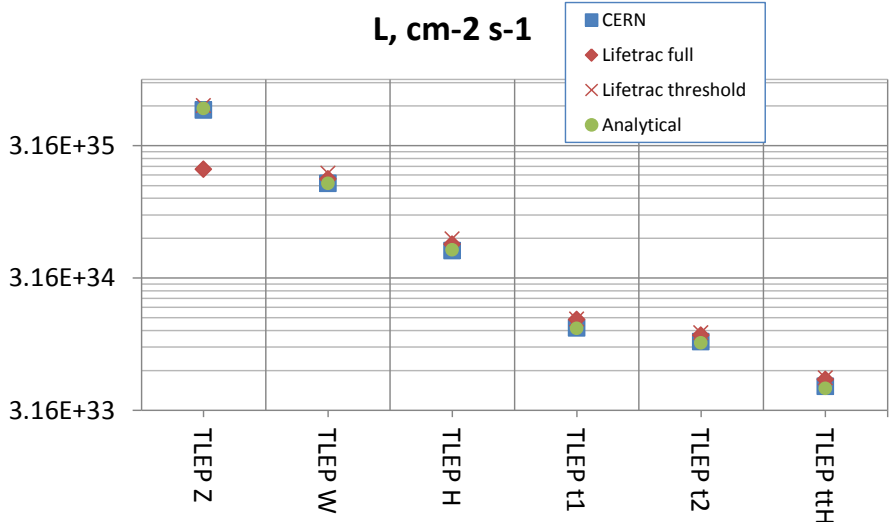}
\caption{Luminosity for different scenarios of TLEP operation. Blue squares are taken from TABLE \ref{tbl:MainParametersCERN}, red diamonds are Lifetrac results with full spectrum of beamstrahlung, red crosses are Lefitrac results if beamstrahlung is considered for emission of photons with energy higher than acceptance, green dots are our analytical calculations.}
\label{fg:Luminosity-0}
\end{figure}

\begin{figure}
\includegraphics[width=5.5cm]{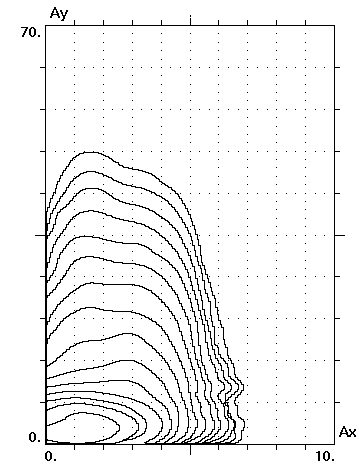}
\includegraphics[width=5.5cm]{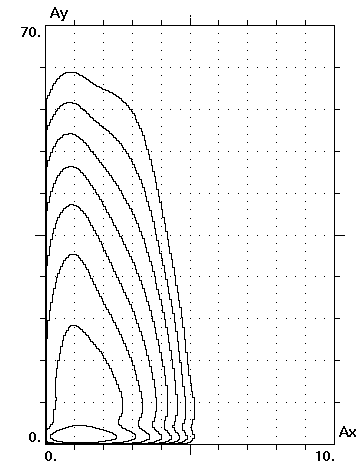}
\caption{Transverse beam distribution in normalized betatron amplitudes for TLEPZ. Left is without beamstrahlung, right is with beamstrahlung. The counter lines are equidistant.}
\label{fg:BeamDistribution-0}
\end{figure}

\begin{figure}
\includegraphics[width=12cm,angle=0,trim=19mm 89mm 19mm 88mm, clip]{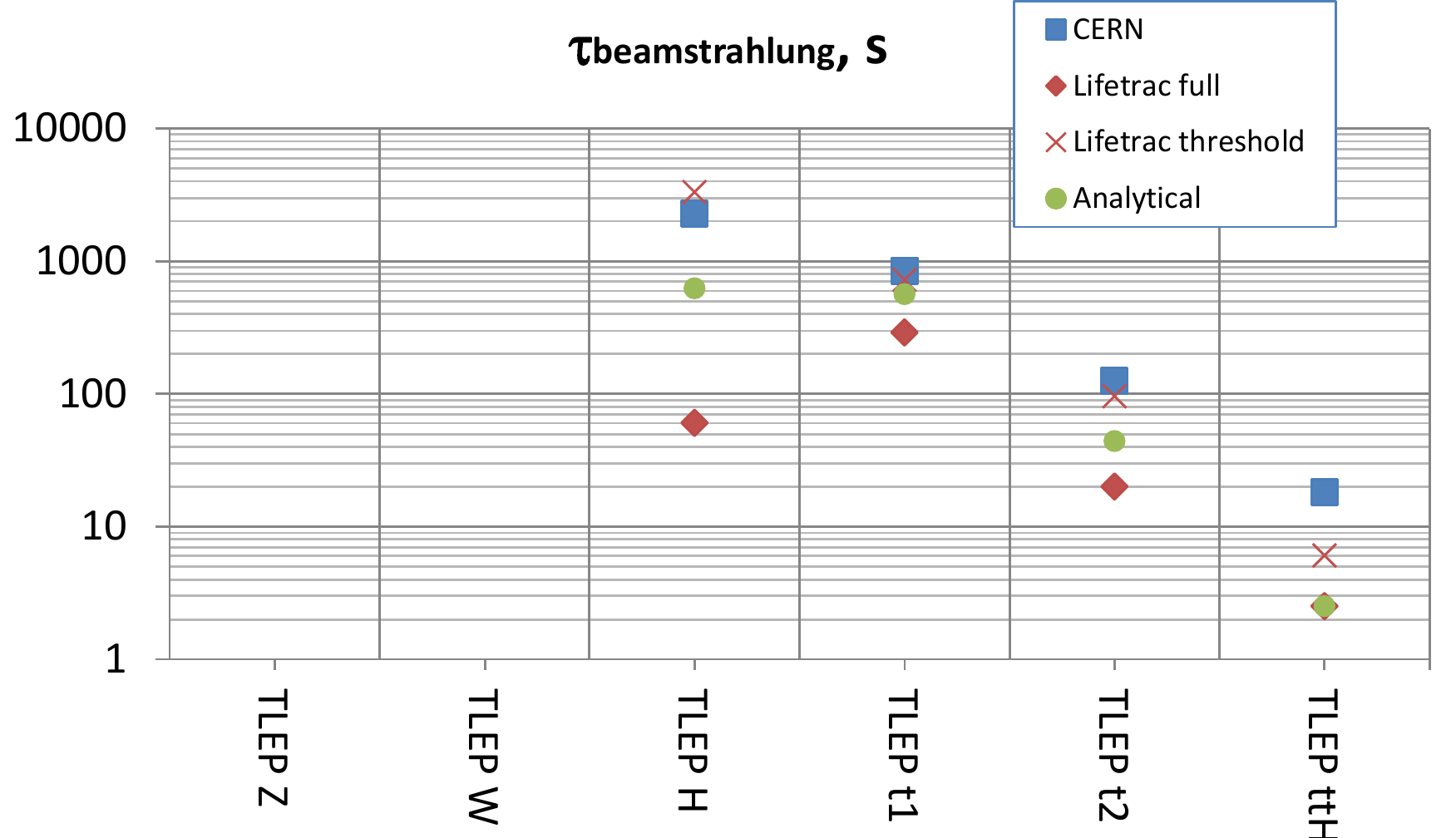}
\caption{Beam lifetime for different scenarios of TLEP operation. Blue squares are taken from TABLE \ref{tbl:MainParametersCERN}, red diamonds are Lifetrac results with full spectrum of beamstrahlung, red crosses are Lefitrac results if beamstrahlung is considered for emission of photons with energy higher than acceptance, green dots are our analytical calculations. Lifetimes for TLEPZ and TLEPW are so large, therefore not plotted. }
\label{fg:LifeTime-0}
\end{figure}

\begin{figure}
\includegraphics[width=12cm,angle=0,trim=19mm 89mm 19mm 88mm, clip]{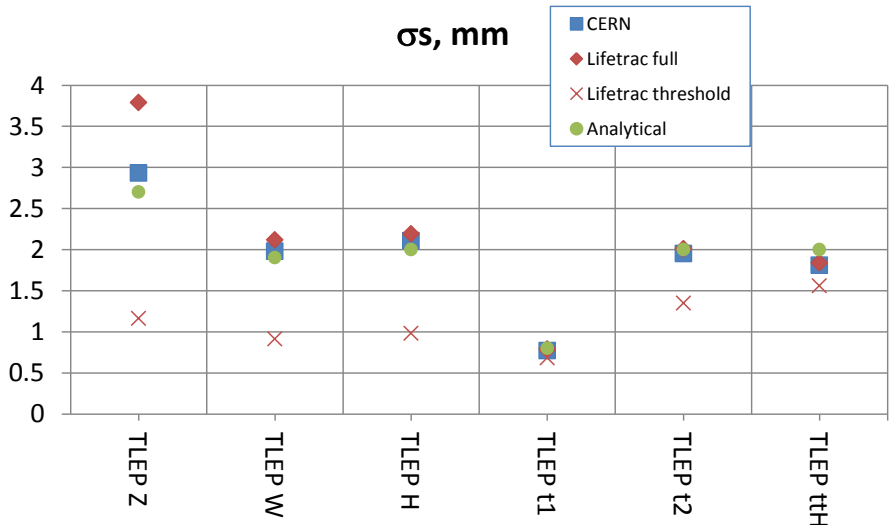}
\caption{Bunch length for different scenarios of TLEP operation. Blue squares are taken from TABLE \ref{tbl:MainParametersCERN}, red diamonds are Lifetrac results with full spectrum of beamstrahlung, red crosses are Lefitrac results if beamstrahlung is considered for emission of photons with energy higher than acceptance, green dots are our analytical calculations.}
\label{fg:BunchLength-0}
\end{figure}

\begin{figure}
\includegraphics[width=12cm,angle=0,trim=19mm 89mm 19mm 88mm, clip]{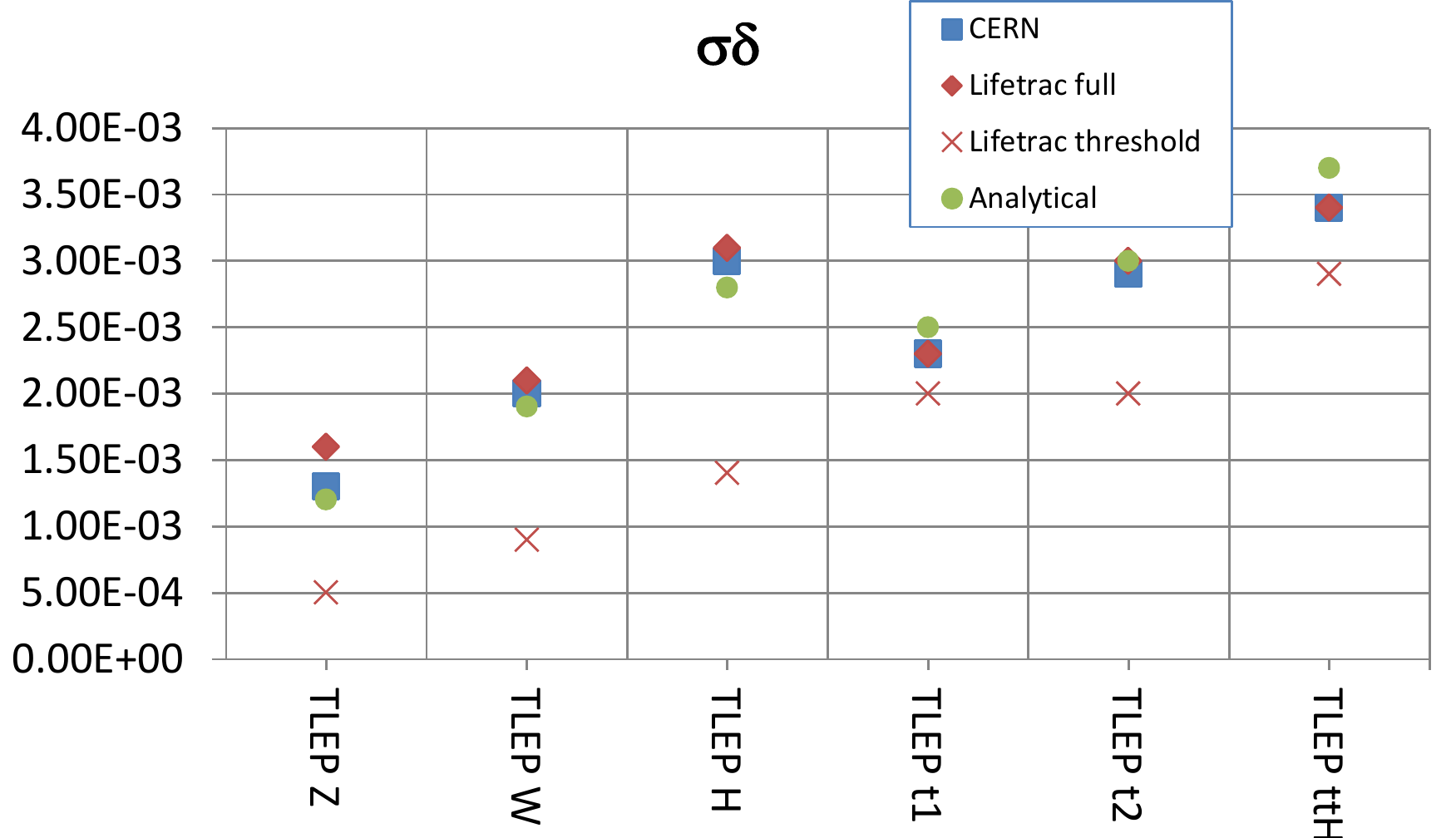}
\caption{Energy spread for different scenarios of TLEP operation. Blue squares are taken from TABLE \ref{tbl:MainParametersCERN}, red diamonds are Lifetrac results with full spectrum of beamstrahlung, red crosses are Lefitrac results if beamstrahlung is considered for emission of photons with energy higher than acceptance, green dots are our analytical calculations.}
\label{fg:EnergySpread-0}
\end{figure}

Initially, we compared our simulation and analytical formula (\ref{eq:LifeTimeOurs}) with the calculations made in CERN. We used  a table of parameters for TLEP given on 24.09.2013 workshop \cite{TLEPparametersCERN}, which are summarized in TABLE \ref{tbl:MainParametersCERN}. Analytical calculations and simulation by Lifetrac and given parameters from TABLE \ref{tbl:MainParametersCERN} of luminosity, beam lifetime, bunch length, energy spread are plotted on Figures \ref{fg:Luminosity-0}, \ref{fg:LifeTime-0}, \ref{fg:BunchLength-0}, \ref{fg:EnergySpread-0} respectively. In all figures CERN stands for CERN calculations from the base table (TABLE \ref{tbl:MainParametersCERN}), Lifetrac full -- quasi strong-strong simulation by Lifetrac with full spectrum of beamstrahlung, Lifetrac threshold -- weak-strong simulation by Lifetrac where only photons with energy higher than energy acceptance are taken into account, meaning that bunch length does not increase (similar to Telnov's approach), analytical -- calculations by (\ref{eq:LifeTimeOurs}) including bunch lengthening (\ref{eq:RadIntegrals-1}, \ref{eq:RadIntegrals-2}).
We understand that computer simulation is not capable of implementing all the effects, but, in the present paper we consider simulation as the most accurate calculation and compare everything against it.
\begin{widetext}
\begin{table}
\caption{Main parameters from 24.09.2013 workshop at CERN \cite{TLEPparametersCERN}}
\label{tbl:MainParametersCERN}
\begin{ruledtabular}
    \begin{tabular}{|l|c|c|c|c|c|c|}                         															    \hline
						& Z			& W			& H		& \multicolumn{2}{c|}{t}		& ttH, ZHH		\\ \hline
$E_{beam}, Gev$				& 45			& 80			& 120		& \multicolumn{2}{c|}{175}	& 250			\\ \hline
Current[mA]					& 1440		& 154			& 29.8	& \multicolumn{2}{c|}{6.7}	& 1.6			\\ \hline
$N_{bunches}$				& 7500		& 3200		& 167		& 160		& 20			& 10			\\ \hline
$N_{particles}[10^{11}]$			& 4.0			& 1.0			& 3.7		& 0.88	& 7.0			& 3.3			\\ \hline
$\varepsilon_x[nm]/\varepsilon_y[pm]$	& 29.2/60		& 3.3/17		& 7.5/15	& 2/2		& 16/16		& 4/4			\\ \hline
$\beta^*_x[m]/\beta^*_y[mm]$	& 0.5/1		& 0.2/1		& 0.5/1	& \multicolumn{2}{c|}{1/1}	& 1/1			\\ \hline
$\sigma_s[mm]$				& 2.93		& 1.98		& 2.11	& 0.77	& 1.95		& 1.81		\\ \hline
$N_{ip}$					& 						 \multicolumn{6}{c|}{4}							\\ \hline
$F_{hg}$ hourglass				& 0.61		& 0.71		& 0.69	& 0.90	& 0.71		& 0.73		\\ \hline
$L/IP[{\scriptscriptstyle 10^{32} cm^{-2}s^{-1}}]$	& 5860 & 1640		& 508		& 132		& 104			& 48			\\ \hline
$\xi_x/IP$					& 0.068		& 0.086		& 0.094	& \multicolumn{2}{c|}{0.057}	& 0.075		\\ \hline
$\xi_y/IP$					& 0.068		& 0.086		& 0.094	& \multicolumn{2}{c|}{0.057}	& 0.075		\\ \hline
$\tau_L$, s					& 5940		& 2280		& 1440	& 1260	& 1560		& 780			\\ \hline
$\tau_{bs}(\eta=2\%)[s]$		& $>\!10^{25}$	& $>\!10^{6}$	& 2280	& 840		& 126			& 18			\\ \hline
$\tau_{||}[turns]$				& 1319		& 242			& 72		& \multicolumn{2}{c|}{23}		& 8			\\ \hline
$f_{s}[kHz]$					& 0.77		& 0.19		& 0.27	& 0.14	& 0.29		& 0.266		\\ \hline
$P_{SR}[MW]$				& 50			& 50			& 50		& 50		& 50			& 50			\\ \hline
    \end{tabular}
\end{ruledtabular}
\end{table}
\end{widetext}
Luminosity calculations by different approaches are consistent except TLEPZ scenario. The difference in the calculated luminosities for TLEPZ scenario is because analytical and probably CERN calculations did not consider beam-beam effects but beamstrahlung. Damping time in TLEPZ scenario is relatively weak, which leads to large bunch lengthening, huge hour-glass, thus, to excitation of synchro-betatron resonances, with a result of a blown-up beam in vertical plane. To illustrate, comparison of transverse beam distributions calculated by Lifetrac without beamstrahlung (left) and with beamstrahlung (right) is shown in FIG.\ref{fg:BeamDistribution-0}.

On the contrary to luminosity calculations agreement, the beam lifetime (FIG.\ref{fg:LifeTime-0}) given by Lifetrac full is consistently smaller than analytical calculations because in analytical calculations particles energy distribution was neglected, however, particle with energy deviation needs to lose different amount of energy in order to be lost. Probability to emit photon with smaller energy is higher, and probability to have a corresponding energy deviation is smaller. The interplay of these probabilities is included in computer simulation, but not in analytical calculations. Additionally, bunch length (Fig.\ref{fg:BunchLength-0}) increases, changing deflecting field and so lifetime;  beam energy spread (Fig.\ref{fg:EnergySpread-0}) becomes larger and energy acceptance of the accelerator shrinks to only $7\div 10$ RMS of energy distribution, thus making particle's loss more probable due to noise excitation. Also, analytical calculations do not include beam sizes dependence on longitudinal position (hour-glass). The Lifetrac threshold simulations of beam lifetime (red crosses on FIG. \ref{fg:LifeTime-0}) correspond well to initial CERN results (blue squares), because simulation used assumption made by V. ~Telnov. Our analytical calculations(green dots on FIG. \ref{fg:LifeTime-0}) are closer to Lifetrac full, especially at TLEPttH.

Bunch length (FIG. \ref{fg:BunchLength-0}) and energy spread (FIG. \ref{fg:EnergySpread-0}) for Lifetrac threshold (red crosses) do not change in calculations because of made assumptions. The discrepancy of bunch length and energy spread between scenarios (red crosses) corresponds to different optics.

Performed comparison shows that accurate simulation gives smaller luminosity at TLEPZ, smaller beam life time in all scenarios. At TLEPttH the beam lifetime is so small (2 sec by Lifetrac full and by our analytics) that given scenario is not feasible.

\section{New set of parameters}
Luminosity for flat beams is given by well known expression
\begin{equation}
\label{eq:Luminosity}
\mathcal{L}=\frac{\gamma}{2er_e}I\frac{\xi_y}{\beta_y}\,,
\end{equation}
where $I$ is a full beam current (limited by synchrotron energy loss), $e$ -- electron charge, $\xi_y$ -- vertical beam-beam tune shift parameter, $\beta_y$ -- minimum beta function at IP. The given value of $\beta_y=1$~mm is already small, further decrease is not reasonable. Hence, luminosity increase is only possible by making $\xi_y$ larger.

Analytical calculations and simulation show that beam-beam effects for TLEP are determined by several factors, quantitative relations between which greatly depend on energy. At high energies (TLEPH and higher) beamstrahlung becomes a main factor which determines beam lifetime. The only way to decrease beamstrahlung influence (\ref{eq:LifeTime-0}) is to increase $\rho$. Since, we do not want to make $\xi_y$ smaller, the only way is to make interaction length $L$ (\ref{eq:InteractionLength-0}) larger (in head-on collision --- by increasing the bunch length). We will assume that bending radius of beamstrahlung is proportional to energy (\ref{eq:BendingRadius-2}) (note that beam sizes and bunch population are changing with energy also).

Another influence of beamstrahlung is increase of the beam energy spread and so the bunch length. Oddly enough, this effect is important at low energies (TLEPZ and TLEPW) but not at high energies. This happens, because relative critical energy $u_c/E_0$ of synchrotron radiation in dipoles raises as $\gamma^2$ (\ref{eq:CriticalEnergy-0}), since bending radius in dipoles does not change, but of beamstrahlung as $\gamma$. Number of photons in both cases is proportional to energy. This is valid for beamstrahlung because interaction length changes with energy in the same manner as bending radius. Thus, the relative input of beamstrahlung in energy spread falls with energy increase.

Apparent paradox of why then at high energies beam lifetime is limited by beamstrahlung is solved by noticing that in spite of faster rise of $u_c$ with energy for conventional synchrotron radiation, beamstrahlung $u_c$ is still significantly higher at all energies, because bending radius in beamstrahlung is at least two orders of magnitude smaller than one of dipoles. Hence, energy of the photons emitted in IP is by two orders of magnitude higher (but amount of them is smaller). Though, beam lifetime is determined by probability to radiate single photon with high energy, which comes from beamstrahlung.

Increasing the bunch length could have a negative effect. When $\beta_y\ll \sigma_s$ (head-on collisions) hour-glass effects decreases the limit of beam-beam tune shift parameter (because of dynamical beta) and makes synchro-betatron resonances stronger (leads to beam blowup). Small damping time counteract the negative influence of synchro-betatron resonances, what happens at high energies (TLEPH and higher). Also, at high energies, the utmost value of beam-beam tune shift parameter is relatively small because it is determined not by conventional beam-beam effects but by beamstrahlung. At low energies, when damping times are larger, bunch lengthening is stronger, hour-glass leads to negative consequences for equilibrium beam distribution in vertical plane (FIG.\ref{fg:BeamDistribution-0}).
CRAB waist collision scheme \cite{CrabWaistPanta} allows to solve this problem. Interaction with large Piwinski parameter allows to make $\beta_y\approx L\ll\sigma_s$ without negative influence of hour-glass. CRAB sextupoles allow to obtain record high beam-beam tune shift parameter $\xi_y$. Yet, at high energies CRAB waist is useless because $\xi_y$ is already limited by beamstrahlung.

On the contrary to vertical plane, CRAB waist does not help to solve problem with synchro-betatron resonances in horizontal plane. The ways to avoid it are either obtain small damping times (what happens at high energies) or provide beam-beam tune shift parameter smaller than synchrotron oscillations frequency (so they are not crossed).

Considering our speculations, we propose the following approach to decide on TLEP parameters at different energies.
At the foundation of the approach is a desire to have the same lattice at all energies and to obtain maximum luminosity with satisfying beam lifetime.
A set of parameters for TLEPH is used as a base, and other scenarios are scaled with respect to energy in emittance, energy spread and energy loss. Bunch length is scaled with energy and adjusted by varying RF amplitude.  We increased synchrotron bunch length to $5$~mm in TLEPH from original $0.98$~mm, kept energy spread the same of $1.4\times10^{-3}$.

At low energies in order to implement CRAB waist collision scheme, we introduced relatively moderate crossing angle  of $2\theta=30$~mrad. The chosen value provides interaction length $L$ approximately equal to vertical beta function ($\beta_y=1$~mm) at TLEPZ and TLEPW. We kept the crossing angle the same for other scenarios so that geometry of interaction region is not changed for all scenarios. Bunch population $N_p$ was chosen to provide horizontal beam-beam tune shift parameter $\xi_x\lesssim 0.03$ (including bunch lengthening by beamstrahlung), which is smaller than synchrotron tune. Vertical emittance was set to give $\xi_y\lesssim 0.2$. Number of bunches was calculated that the total power loss does not exceed 50~Mw.

At high energies, bunch length becomes large, therefore crossing angle helps to reduce length of interaction area making hour-glass reasonable. However, the main concern is not to minimize hour-glass effect but to increase beam lifetime. Therefore, number of particles was chosen to provide good beam lifetime. Number of bunches and vertical emittance were set to obtain maximum luminosity and not to exceed 50 MW of total power loss. Piwinski parameter at these scenarios is not small ($\phi>2.5$), therefore CRAB sextupoles will decrease beam blowup because of beam-beam effects.

For all scenarios, crossing angle helps to facilitate separation of the bunches and accommodation of the final focus elements.

\begin{figure}
\includegraphics[width=12cm,angle=0,trim=19mm 89.0mm 19mm 88mm, clip]{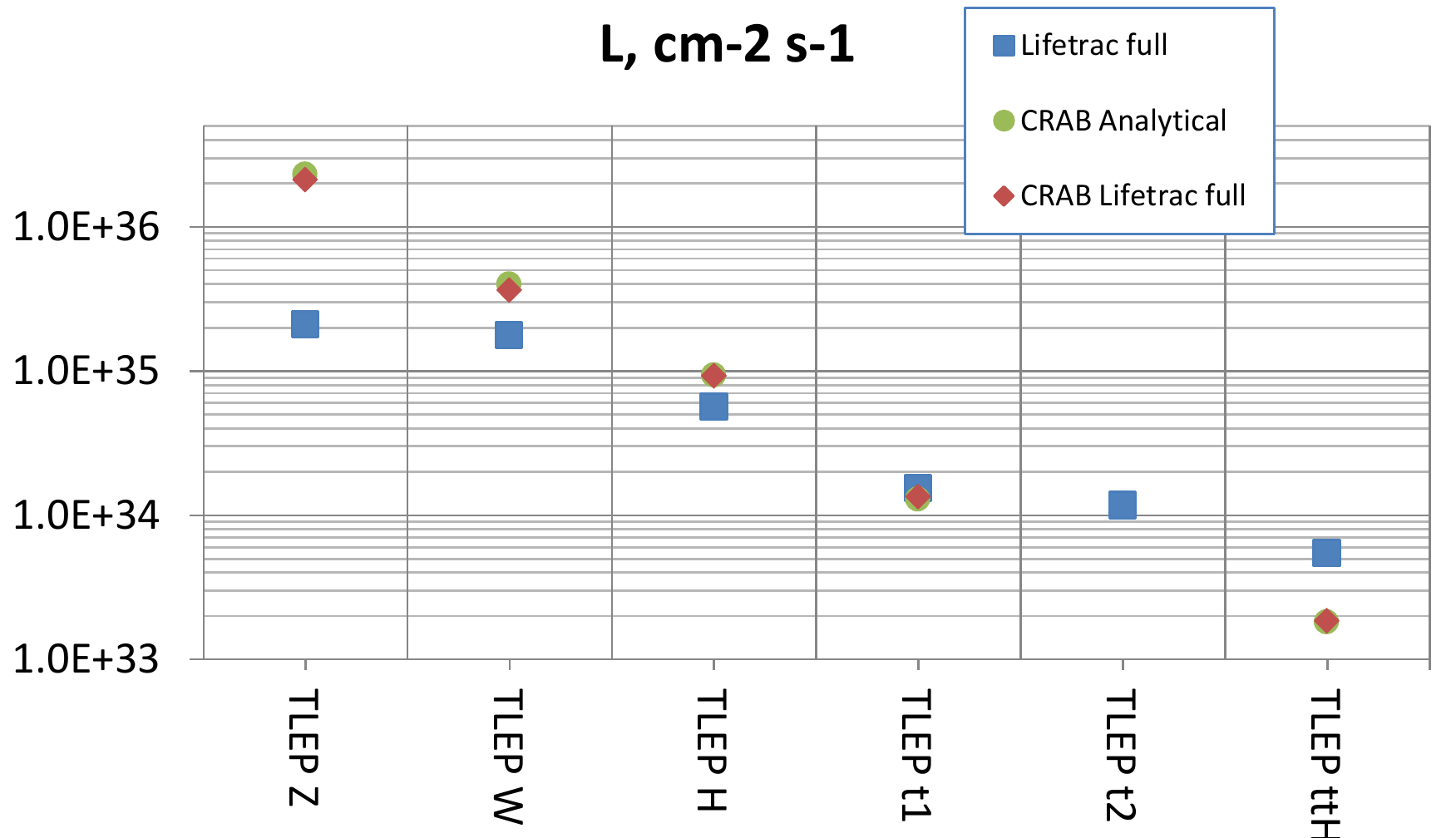}
\caption{Luminosity for different scenarios of TLEP operation. Blue squares are Lifetrac calculations with full spectrum of beamstrahlung for TABLE \ref{tbl:MainParametersCERN}, red diamonds are Lifetrac results with full spectrum of beamstrahlung for the new set of parameters, green dots are analytical calculations for the new set of parameters.}
\label{fg:Luminosity-1}
\end{figure}

\begin{figure}
\includegraphics[width=12cm,angle=0,trim=19mm 89mm 19mm 88mm, clip]{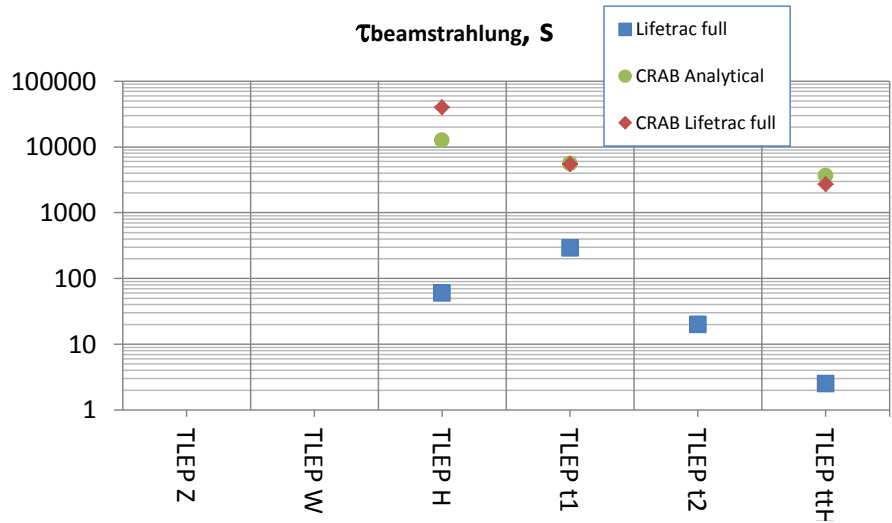}
\caption{Beam lifetime for different scenarios of TLEP operation. Blue squares are Lifetrac calculations with full spectrum of beamstrahlung for TABLE \ref{tbl:MainParametersCERN}, red diamonds are Lifetrac results with full spectrum of beamstrahlung for the new set of parameters, green dots are analytical calculations for the new set of parameters.}
\label{fg:LifeTime-1}
\end{figure}

The new set of parameters is given in TABLE \ref{tbl:ParametersCRAB-1}.
\begin{table}
\caption{A new set of parameters with crossing angle and CRAB waist}
\label{tbl:ParametersCRAB-1}
\begin{ruledtabular}
    \begin{tabular}{|l|c|c|c|c|c|}                         											    \hline
						& Z			& W				& H			& t			& ttH, ZHH		\\ \hline
$\Pi[km]$					&  \multicolumn{5}{c|}{100}												\\ \hline
$2 \theta[mrad]$				&  \multicolumn{5}{c|}{30}												\\ \hline
Current[mA]					& 1431		& 142				& 29			& 6.3			& 1.4			\\ \hline
$N_{bunches}$				& 29791		& 739				& 127			& 33			& 6			\\ \hline
$N_{particles}[10^{11}]$			& 1			& 4				& 4.7			& 4			& 5			\\ \hline
$\varepsilon_x[nm]/\varepsilon_y[pm]$	& 0.14/1		& 0.44/2			& 1/2			& 2.1/4.25		& 4.34/8.68		\\ \hline
$\beta^*_x[m]/ \beta^*_y[m]$		& \multicolumn{5}{c|}{0.5/0.001}											\\ \hline
$F_{RF}[MHz]$				&  \multicolumn{5}{c|}{$300$}												\\ \hline
$V_{RF}[GV]$				& 0.54		& 1.35			& 3.6			& 11.4		& 34.2		\\ \hline
$\nu_{syn}$					& 0.062		& 0.072			& 0.092		& 0.124		& 0.124		\\ \hline
$\delta_{RF\, bucket}[\%]$		& 5.9			& 5.9				& 6			& 6.1			& 2.6			\\ \hline
mom. comp. $\alpha$			&  \multicolumn{5}{c|}{$2\cdot 10^{-5}$}										\\ \hline
$\sigma_{s, syn}[mm]$			& 2.7			& 4.1				& 4.9			& 5.3			& 7.5			\\ \hline
$\sigma_{\delta, syn}[10^{-3}]$		& 0.5			& 0.9				& 1.4			& 2			& 2.9			\\ \hline
$\sigma_s[mm]$				& 5.9			& 9.1				& 8.2			& 6.6			& 8			\\ \hline
$\sigma_{\delta}[10^{-3}]$		& 1.2			& 2.1				& 2.4			& 2.6			& 3.1			\\ \hline
$F_{hg}$ hourglass				& 0.94		& 0.86			& 0.78		& 0.7			& 0.61		\\ \hline
$L/IP[{\scriptscriptstyle 10^{32} cm^{-2}s^{-1}}]$	& 22971	& 3977		& 933			& 129			& 18			\\ \hline
$\xi_x/IP$					& 0.032		& 0.031			& 0.029		& 0.024		& 0.014		\\ \hline
$\xi_y/IP$					& 0.175		& 0.187			& 0.16		& 0.077		& 0.038		\\ \hline
$\tau_L[s]$					& 2294		& 1315			& 1132		& 1814		& 2942		\\ \hline
$\tau_{bs}(\eta=2\%)[s]$		& $>\!10^{19}$	& $>\!10^{6}$		& 12468		& 5551		& 3636		\\ \hline
$\tau_{||}[turns]$				& 1338		& 238				& 70			& 22			& 7			\\ \hline
$U_{loss, SR}[GeV/turn]$			& 0.03		& 0.3				& 1.7			& 7.7			& 32			\\ \hline
$P_{SR}[MW]$				& 50			& 50				& 50			& 49.1		& 46.3		\\ \hline
    \end{tabular}
\end{ruledtabular}
\end{table}
Our proposal compared against Lifetrac full for original CERN set of parameters (Figures \ref{fg:Luminosity-1}, \ref{fg:LifeTime-1}) gives 10 times higher luminosity at TLEPZ,  2 times higher luminosity at TLEPW and 1.6 times higher at TLEPH, almost the same luminosity for TLEPt and 20 times higher beamstrahlung lifetime, 2.9 times smaller luminosity at TLEPtth but realistic beamstrahlung lifetime 3636 sec against 2 sec. On the figures Lifetrac full stands for simulation of the original table of parameters (TABLE \ref{tbl:MainParametersCERN}, blue squares), CRAB analytical and CRAB Lifetrac full are calculations (green dots) and simulations (red diamonds) respectively, for the new set (TABLE \ref{tbl:ParametersCRAB-1}).

\begin{figure}
\includegraphics[width=12cm,angle=0,trim=19mm 89mm 19mm 88mm, clip]{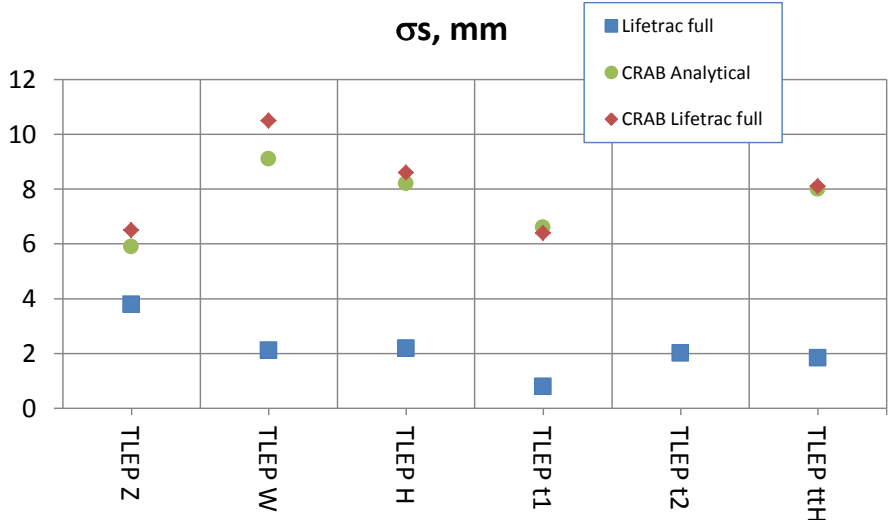}
\caption{Bunch length for different scenarios of TLEP operation. Blue squares are Lifetrac calculations with full spectrum of beamstrahlung for TABLE \ref{tbl:MainParametersCERN}, red diamonds are Lifetrac results with full spectrum of beamstrahlung for the new set of parameters, green dots are analytical calculations for the new set of parameters.}
\label{fg:BunchLength-1}
\end{figure}

\begin{figure}
\includegraphics[width=12cm,angle=0,trim=19mm 89mm 19mm 88mm, clip]{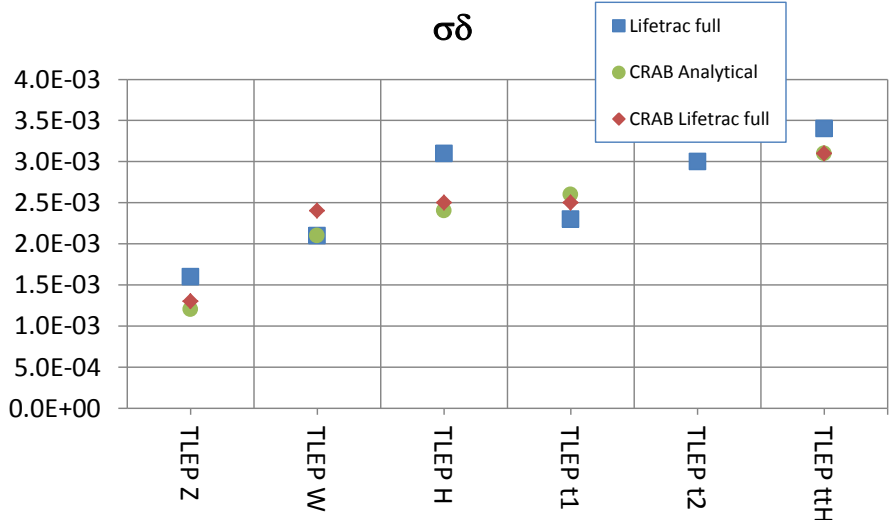}
\caption{Energy spread for different scenarios of TLEP operation. Blue squares are Lifetrac calculations with full spectrum of beamstrahlung for TABLE \ref{tbl:MainParametersCERN}, red diamonds are Lifetrac results with full spectrum of beamstrahlung for the new set of parameters, green dots are analytical calculations for the new set of parameters.}
\label{fg:EnergySpread-1}
\end{figure}

The bunch length and energy spread are shown on FIG. \ref{fg:BunchLength-1} and \ref{fg:EnergySpread-1}. Analytical calculations correspond well Lifetrac simulation at TLEPH, TLEPt and TLEPttH. Discrepancy at TLEPZ and TLEPW happens because analytical calculations do not consider horizontal and vertical emittance increase owing to beam beam effects.

\section{Conclusion}
We have considered different aspects of the beamstrahlung influence on the parameters of the high-energy high-luminosity e+e- storage ring collider TLEP operating in the energy range from Z-pole up to the $t\bar{t}$ threshold. Consideration only of the single beamstrahlung is not sufficient to optimize the machine specifications in the entire energy range. Energy loss due to the multiple beamstrahlung increases bunch length and energy spread, modifies  probability to emit photons. Particle with energy deviation might emit photon with smaller energy but higher probability to be outside of energy acceptance. Thus, the beam lifetime could be several times smaller than that predicted by the single beamstrahlung formalism.

Accurate consideration of beamstrahlung influence requires quasi strong-strong or strong-strong simulation with damping and noise excitation. Analytical approach does not consider all the effects, however gives sufficient estimation.
The new set of parameters enhances performance of TLEP.

\begin{acknowledgments}
We would like to thank V.Telnov, M. Koratzinos and F. Zimmermann for fruitful discussions.
The work is supported by Russian Ministry of Education and Science.

\end{acknowledgments}

\bibliography{References.bib}

\end{document}